\documentclass[12pt]{article}
\usepackage{graphicx}
\usepackage{slashed}
\usepackage{amsmath, amssymb}

\newcommand\pubnumber{}
\newcommand\pubdate{\today}

\def\institute{Department of Physics, University of Cincinnati, Cincinnati, Ohio 45221, USA}
\def\authemail{\footnote{Contact: szewcml@ucmail.uc.edu}}

\def\Title#1{\begin{center} {\Large #1 } \end{center}}
\def\Author#1{\begin{center}{ \sc #1} \end{center}}
\def\Address#1{\begin{center}{ \it #1} \end{center}}

\newcommand\pubblock{\rightline{\begin{tabular}{l} \pubnumber\\
         \pubdate  \end{tabular}}}
\newenvironment{Abstract}{\begin{quotation}  }{\end{quotation}}
\newenvironment{Presented}{\begin{quotation} \begin{center} 
             PRESENTED AT\end{center}\bigskip 
      \begin{center}\begin{large}}{\end{large}\end{center} \end{quotation}}
\def\Acknowledgements{\bigskip  \bigskip \begin{center} \begin{large}
             \bf ACKNOWLEDGEMENTS \end{large}\end{center}}

\def\beq{\begin{equation}}
\def\eeq#1{\label{#1}\end{equation}}
\def\eeqn{\end{equation}}

\def\beqa{\begin{eqnarray}}
\def\eeqa#1{\label{#1}\end{eqnarray}}
\def\eeqan{\end{eqnarray}}

\let\bar=\overbar

\def\Dslash{\not{\hbox{\kern-4pt $D$}}}
\def\dslash{\not{\hbox{\kern-2pt $\del$}}}

\def\msb{{\bar{\ssstyle M \kern -1pt S}}}

\begin{document}
\begin{titlepage}
\pubblock

\vfill
\Title{Accessing CKM suppressed top decays at the LHC}
\vfill
\Author{ Manuel Szewc\authemail}
\Address{\institute}
\vfill
\begin{Abstract}

We present an strategy for measuring the off-diagonal elements of the third row of CKM matrix $|V_{tq}|$ through the branching fractions of top quark decays $t\to q W$, where $q$ is a light quark jet. This strategy is an extension of existing measurements, with the improvement rooted in the use of orthogonal $b$- and $q$-taggers that add a new observable, the number of light-quark-tagged jets, to the already commonly used observable, the fraction of $b$-tagged jets in an event. Careful inclusion of the additional complementary observable significantly increases the expected statistical power of the analysis, with the possibility of excluding a null $|V_{td}|^2+|V_{ts}|^2$ at $95\%$ C.L. at the HL-LHC.
\end{Abstract}
\vfill
\begin{Presented}
$16^\mathrm{th}$ International Workshop on Top Quark Physics\\
(Top2023), 24--29 September, 2023
\end{Presented}
\vfill
\end{titlepage}
\def\thefootnote{\fnsymbol{footnote}}
\setcounter{footnote}{0}

\section{The need for direct $|V_{tq}|$ measurements}

The CKM matrix encodes the flavor structure of the electroweak charged currents in the Standard Model (SM). The precise measurement of the CKM allows for a powerful precision test of the SM with potential sensitivity to Beyond the Standard Model (BSM) effects. This is specially true for the third row $|V_{tq}|$ where the strong hierarchy between its off-diagonal elements and $|V_{tb}|$ makes them difficult to measure but particularly sensitive to possible BSM effects. Indirect constraints on $|V_{tq}|$, such as radiative $B$ meson decays and neutral $B_{s,d}$ meson oscillations, can be compared with direct measurements, either from productions of on-shell top quarks at the LHC and their decays~\cite{Khachatryan:2014nda} or through $t$-channel single top production~\cite{CMS:2020vac}. Although alternative ways of directly measuring $|V_{tq}|$ have been proposed, either using $tW$ associated production~\cite{Alvarez:2017ybk}, or  by $s$-tagging top-quark decay products~\cite{Zeissner:2021nhh}, these approaches currently suffer from low statistics due to the smallness of $\sqrt{|V_{td}|^2+|V_{ts}|^2}$ and are thus not expected to match the precision of the SM prediction from the CKM global fits.
Currently, the best direct bound on $\sqrt{|V_{td}|^2+|V_{ts}|^2}$ is derived from the measurements of $b$-jet fractions in $t \to W j$ top decays reported in Ref.~\cite{Khachatryan:2014nda}
\begin{equation}
\label{eq:Rb:def}
\mathcal{R}_b\equiv\frac{\mathcal{B}(t\to b W)}{\sum_{j=d,s,b} \mathcal{B}(t \to j W)} > 0.955~ @ ~95\% ~\rm C.L.\,,
\end{equation}
which can be interpreted as $\sqrt{|V_{td}|^2+|V_{ts}|^2} < 0.217 |V_{tb}|$. In Ref.~\cite{Faroughy:2022dyq}, we propose an extension of Ref.~\cite{Khachatryan:2014nda} that targets the non-diagonal elements of the third row without sacrificing the statistical power of dileptonic $t\bar t$.

\section{Better statistical power through $q$-tagging}

\begin{figure}[!h!tbp]
\centering
\includegraphics[width=0.5\linewidth]{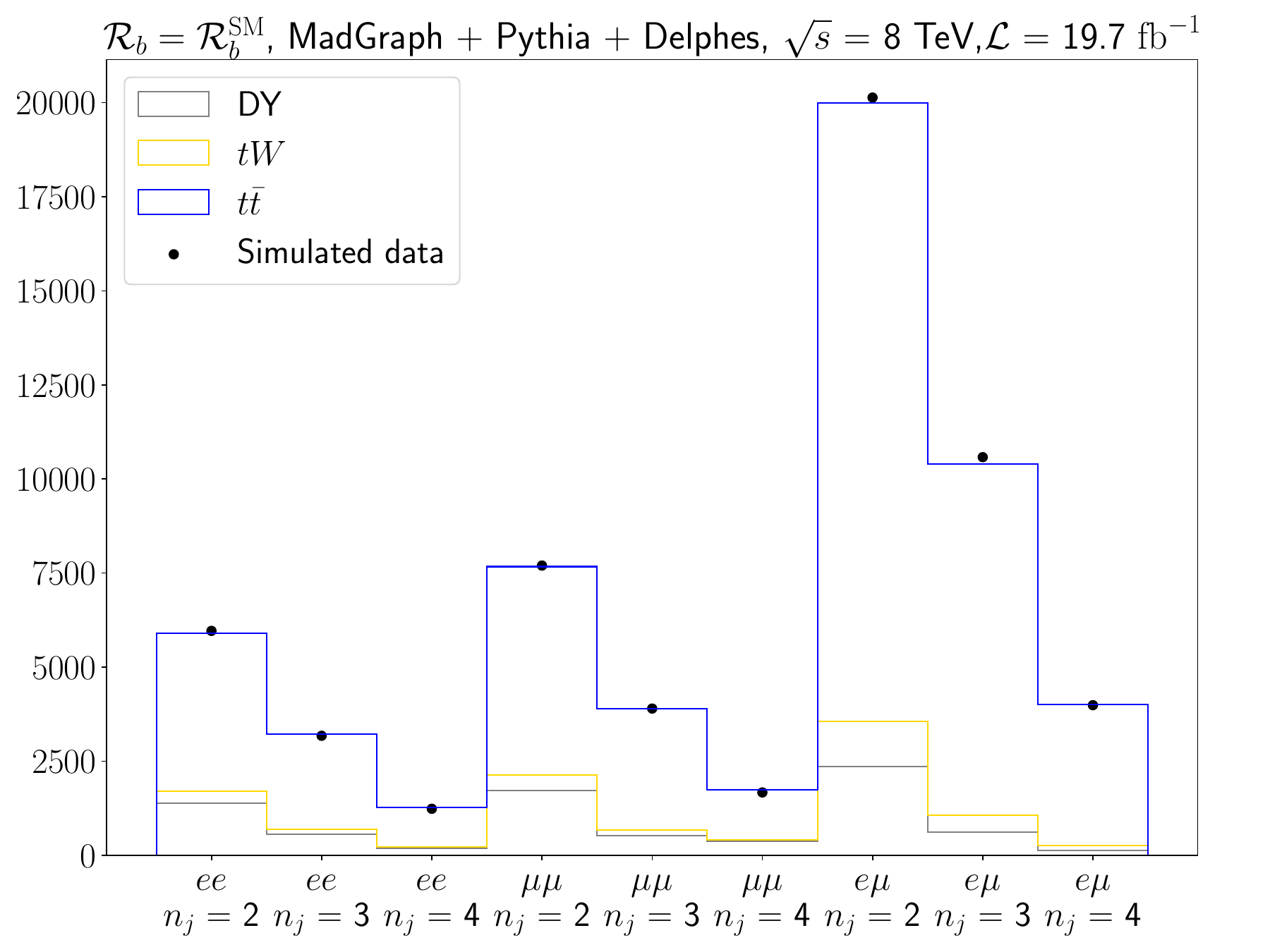}
\caption{The expected number of events for $\mathcal{R}_{b}=\mathcal{R}^{\mathrm{SM}}_{b}$ in each of the $\{\ell\ell',n_j\}$ categories, for $t\bar t$ signal (blue), as well as the $tW$ (yellow) and Drell-Yan (green)  backgrounds, shown in a stacked histogram form. An example of simulated data is denoted with black dots.}
\label{fig:nj_spectra}
\end{figure}

As done originally in Ref.~\cite{Khachatryan:2014nda}, we propose to use the dileptonic $t\bar t$ signal to constrain $\mathcal{R}_b$. To do so, the $pp\to t\bar t$ events (including background events) are first split into different categories, labelled by $\{\ell \ell', n_j\}$, where $\ell \ell'$ are the flavors of the two final state leptons,  $\ell \ell' = e^{+}e^{-},\mu^{+}\mu^{-},e^{\pm}\mu^{\mp}$,  while  $n_j=2,3,4,$ is the number of jets in the event. For our analysis, we compute the expected number of events in each of the $\{\ell\ell',n_j\}$ categories using the {\tt Madgraph}~\cite{Alwall:2014hca}, {\tt Pythia}~\cite{Sjostrand:2014zea}, {\tt Delphes}~\cite{deFavereau:2013fsa} simulation pipeline for $\sqrt{s}=8$\,TeV LHC collision energy and considering the three main production channels: $t\bar t$ with up to two additional jets, $tW$ with no additional jets, and Drell-Yan with up to two additional jets. The obtained expected number of events per category are shown in Figure~\ref{fig:nj_spectra}.

In Ref.~\cite{Faroughy:2022dyq} we use two orthogonal $b$- and $q$-taggers to further divide events according to how many of the $n_{j}$ jets are either $b$-tagged or $q$-tagged, with each event populating different $\{\ell \ell', n_j, n_b, n_q\}$ bins. The expected number of events in the $\{\ell \ell', n_j, n_b,n_q\}$ bin is then expressed in terms of a probabilistic model that depends on $\mathcal{R}_b$,  the parameter we are interested in, as well as on a number of nuisance parameters, $\theta_i$, discussed extensively in Refs.~\cite{Khachatryan:2014nda} and~\cite{Faroughy:2022dyq}. We constraint $\mathcal{R}_b$ by comparing the expected number of events, $ \bar{N}_{\ell \ell'}(n_b,n_q|n_j)$, with the observed number of events in the $\{\ell \ell', n_j, n_b,n_q\}$ bin, $N_{\ell \ell'}(n_b,n_q|n_j)$, through the likelihood function. The new model, which is detailed in Ref.~\cite{Faroughy:2022dyq}, can be reduced to the model introduced in Ref.~\cite{Khachatryan:2014nda} simply by marginalizing over $n_{q}$.

To validate our analysis, we build simple $b$- and $q$-taggers to tag our simulated jets. We consider a simple secondary vertex-based $b$-tagger~\cite{CMS:2012feb} and we build a $q$-tagger combining the same the secondary vertex-information, to render it orthogonal to the $b$-tagger, with the constituent multiplicity of the jet to discriminate between quarks and gluons~\cite{Gallicchio:2012ez}. We observe that the resulting $68\%$ confidence intervals improve upon the strategy introduced in Ref.~\cite{Khachatryan:2014nda}; although the improved fit sensitivity to $\mathcal{R}_{b}$ is still not large enough to separate $\mathcal{R}^{\mathrm{SM}}_{b}$ from $\mathcal{R}_{b}=1$ and thus from $|V_{td}|^{2}+|V_{ts}|^{2}=0$ at $68\%$ C.L.. More importantly, even using suboptimal taggers, with quite likely inflated systematic uncertainties, and without incorporating the full $p_{T}$ dependence of the tagging  efficiencies, the model is flexible enough to represent the true distributions of the measured observables $\{n_b, n_q\}$ and provide an unbiased estimator for $\mathcal{R}_{b}$. We conclude that the probabilistic model is an apt description of the relevant physics for $\mathcal{R}_{b}$ measurements.

\section{Incorporating state-of-the-art taggers}

Having validated our probabilistic model on simulated data and established the possible increase in statistical power through the addition of the $q$-tagger, we suggest an improved analysis strategy, which we work out in this section. To project the power of said strategy, we assume the probabilistic model to be a faithful representation of the relevant physics related to $\mathcal{R}_{b}$ measurements and we sample pseudo-data consisting of $\{\ell,\ell',n_j,n_b,n_q\}$ counts using the probabilistic model, the simulated $\{\ell,\ell',n_j\}$ event yields and specific choices for the parameters. We then perform our statistical analyses on this pseudo-data. The use of pseudo-experiments is motivated by two main factors: we can incorporate any tagger we want as long as we can access the reported efficiencies and associated uncertainties, and we can cheaply explore different choices of $\mathcal{R}_{b}$ and other parameters of interest.

The proposed strategy is very simple: we make use of state-of-the-art $b$-taggers and quark/gluon taggers to define complementary orthogonal regions. To ensure orthogonality, we choose two Working Points (WP) of the former to define a $b_{1}$-tagger and an anti-$b_2$-tagger. This anti-$b_2$-tagger is combined with the quark/gluon tagger to define an improved version of a $q$-tagger. To do this, we take advantage of the fact that state of the art $b$-taggers provide a spectrum of working points, each of which defines $\{n_{b},n_{q}\}$ bins of varying sample purity.  To obtain $\{n_{b},n_{q}\}$ bins of varying purity with the two working points, WP1 and WP2,  we first apply the $b_{1}$-tagger, giving $n_b$  $b$-tagged jets for each event and then combine the anti-$b_{2}$-tagger with the quark/gluon tagger to obtain the $n_q$ $q$-tagged jets for each event. The definition of $n_q$ bins is more involved because we want to obtain a high-purity $q$-tagged jet sample (with almost no $b$ quarks), starting with the jets that were rejected by the $b_{1}$-tagger. This cannot be achieved by simply applying a quark/gluon-tagger to these jets, since even the state-of-the-art quark/gluon-taggers usually still group together the $b$-quarks and $q$-quarks. However, we can combine the quark/gluon-tagger with an anti-$b$-tagger (using WP2), which gives a $q$-tagger that is orthogonal to the $b$-tagger (from WP1). That is, the $q$-tagged jets belong to the intersection of anti-$b_{2}$-tagged and quark/gluon-tagged jets, so that the efficiency of the $q$-tagger is $\epsilon^{q}=\epsilon^{\{\rm{anti}-b_{2}\}\cap \{q/g\}}$.  For the anti-$b$-tagger we use the WP2 of the $b$-tagger, since the anti-$b_{2}$ tagger is more powerful in rejecting $b$ quark jets than the anti-$b_{1}$-tagger.  The $b$- and $q$-taggers defined in this way select non-overlapping $q$- and $b$-tagged jet fractions, i.e., they are orthogonal.

In our numerical analysis we consider two state-of-the art $b$-taggers: for $\sqrt{s} = 8 \mathrm{TeV}$ collision energy events we use the CSV tagger~\cite{CMS:2012feb}, and for  $\sqrt{s} = 13 \mathrm{TeV}$ the cMVAv2 tagger~\cite{CMS:2017wtu}. Additionally, we assume for convenience that the quark/gluon-tagger always selects a fixed subset of the anti-$b_{2}$-tagged jets and thus $\epsilon^{q}=\epsilon^{q/g}$. This simplifies our analysis, but it does mean that the quark/gluon tagger working point is modified for each choice of WP2.  That $\epsilon^{q}$ are varied does have a practical advantage, since we can explore different WP regimes. For low $\epsilon^{b_{2}}$ the resulting $q$-tagger will lead to high sample size at the expense of sample purity, while for high $\epsilon^{b_{2}}$ sample size decreases considerably as sample purity increases. For the experimental analysis the choices of $\epsilon^{q}$ could be further optimized, however, we  expect the qualitative conclusions about the added statistical power provided by $\{n_{q}\}$ binning to be robust. With this set-up, the partitioning procedure is as follows: we first $b_{1}$-tag the objects, obtaining $n_{b}$ $b_{1}$-tagged jets. All remaining jets are subjected to the anti-$b_{2}$-tagger. Finally, we apply a quark/gluon-tagger to all the jets that are anti-$b_{2}$-tagged, obtaining $n_{q}$ $q$-tagged jets. 

\section{Projected sensitivity}

To compare Working Points in a simple way, we reframe the problem as a signal discovery where we have a null hypothesis $\mathcal{R}_{b} = 1$ and an alternative hypothesis $\mathcal{R}_{b} = \mathcal{R}^{\mathrm{SM}}_{b}$. To obtain the expected discovery significance, we follow Ref.~\cite{Cowan:2010js} and build a test statistic $q_{1}$ to reject the $\mathcal{R}_{b}=1$ hypothesis, i.e., the hypothesis that $|V_{td}|^2+|V_{ts}|^2 = 0$, assuming the true value of $\mathcal{R}_b$ is the SM one, $\mathcal{R}_{b}=\mathcal{R}^{\mathrm{SM}}_{b}$. The expected median significance $Z_{1}$ is then given by the median of $\sqrt{q_{1}}$. To reduce the computational costs we make use of the Asimov approximation~\cite{Cowan:2010js} and assume that $Z_1$ is given by $Z_1=\sqrt{q_{1,A}}$, where the value $q_{1,A}$ is computed using the Asimov dataset, i.e., a dataset with each bin yield is equal to the expected rate given by the probabilistic model and taking $\mathcal{R}_b=\mathcal{R}^{\mathrm{SM}}_{b}$ and with the nuisance parameters $\theta_{i}$ set to their central values. We have verified explicitly the validity of the Asimov approximation for several WPs by performing a set of pseudo-experiments and verifying that the distribution of the test statistics $q_1$ approaches its asymptotic limit, a non-central chi-squared distribution, with the median approximated well by $q_{1,A}$.

\begin{figure}[!h!tbp]
\begin{center}
 \begin{tabular}{cc}
 \hspace*{-5mm}
 \includegraphics[width=0.5\linewidth]{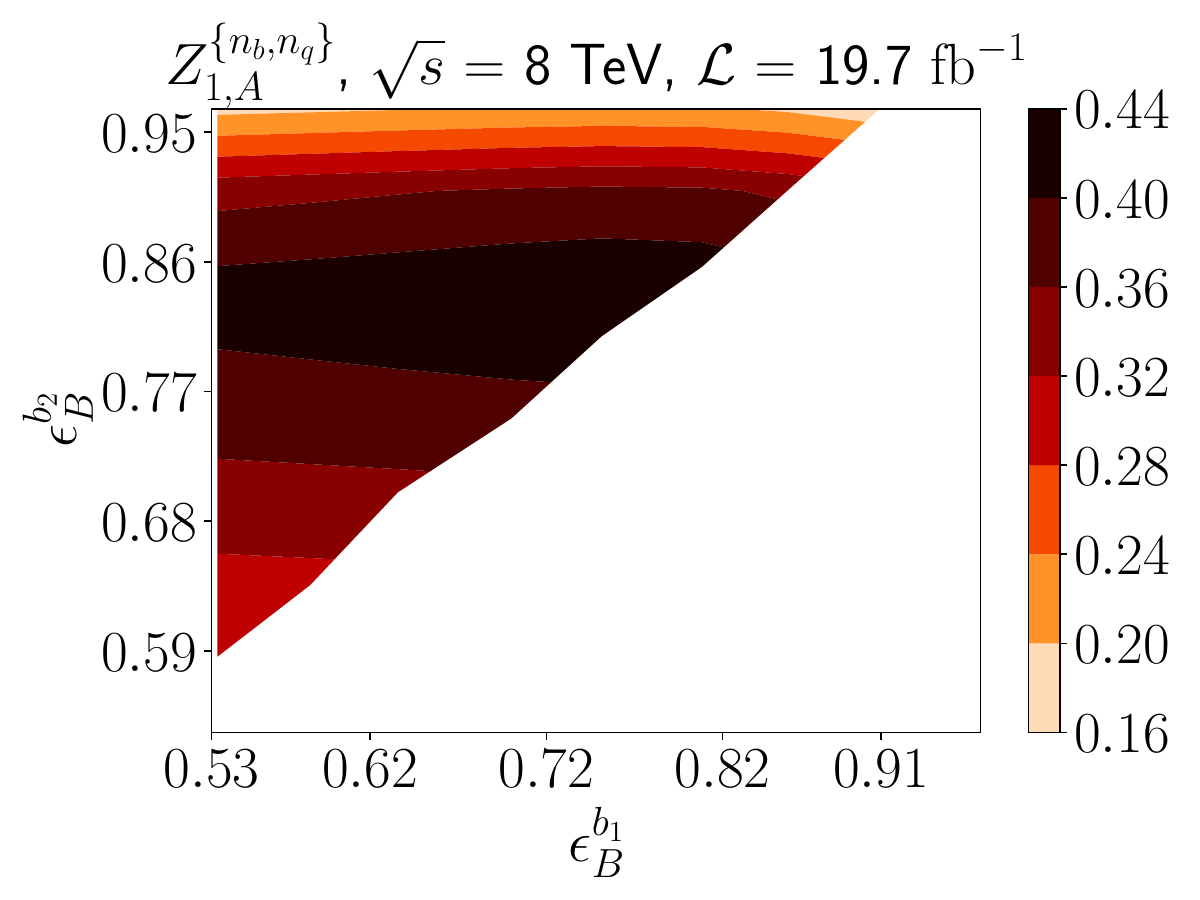}
 \includegraphics[width=0.5\linewidth]{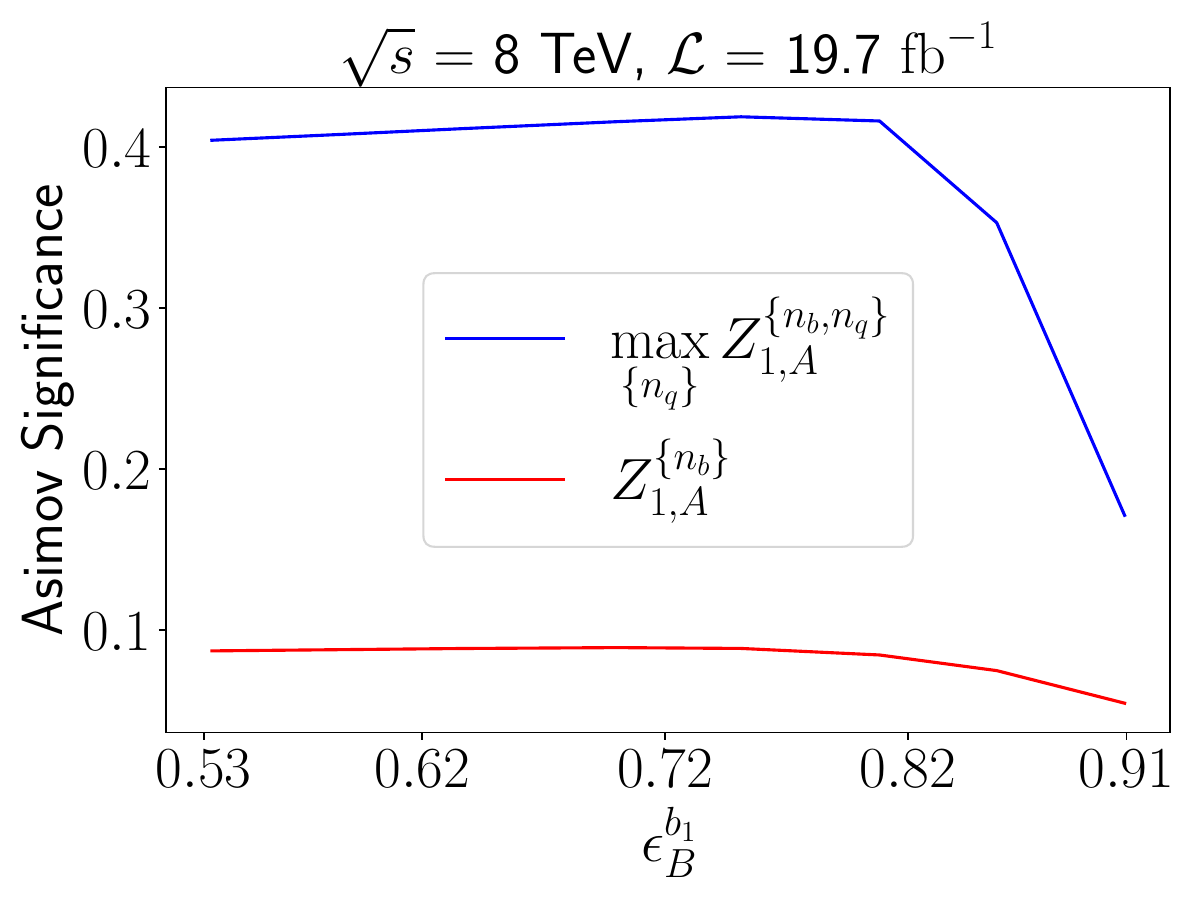}
 \\
 \hspace*{-5mm}
 \includegraphics[width=0.5\linewidth]{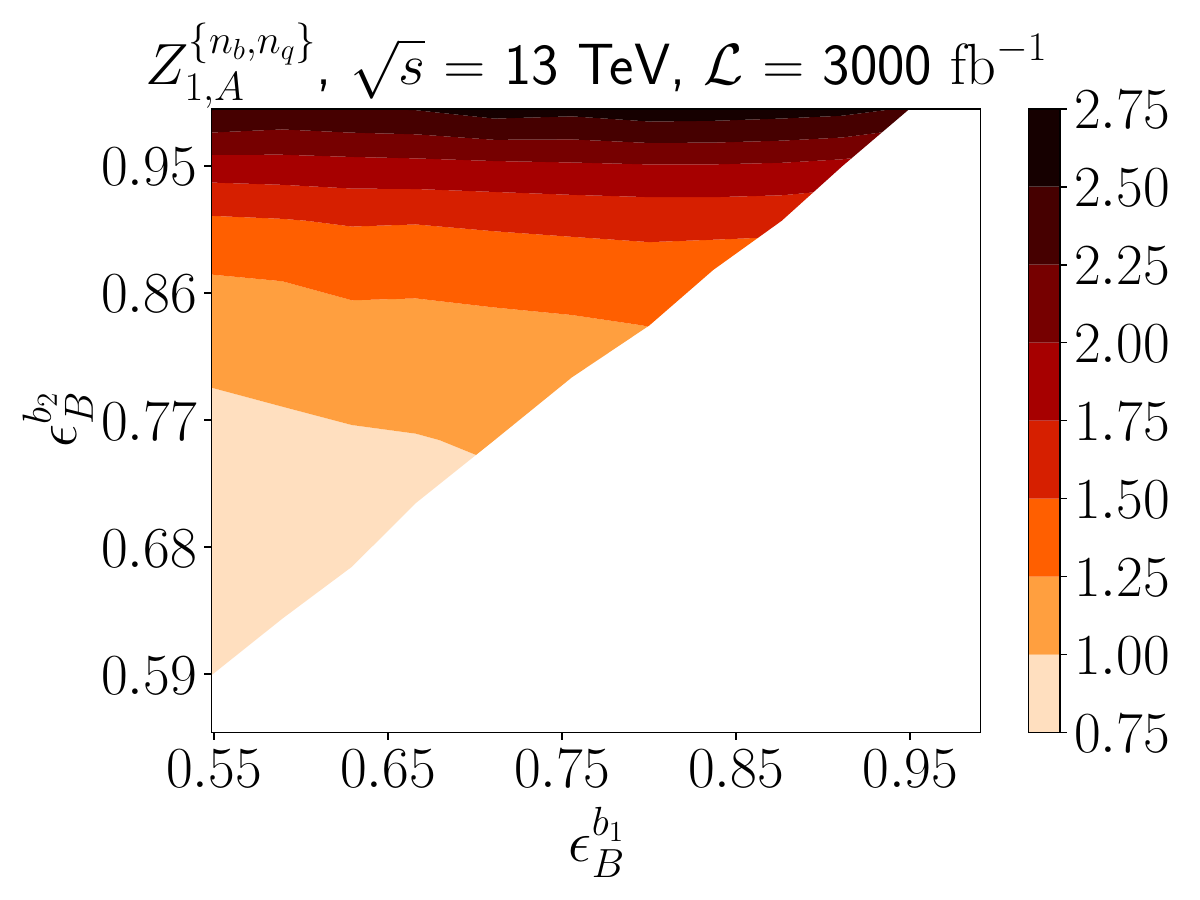}
 \includegraphics[width=0.5\linewidth]{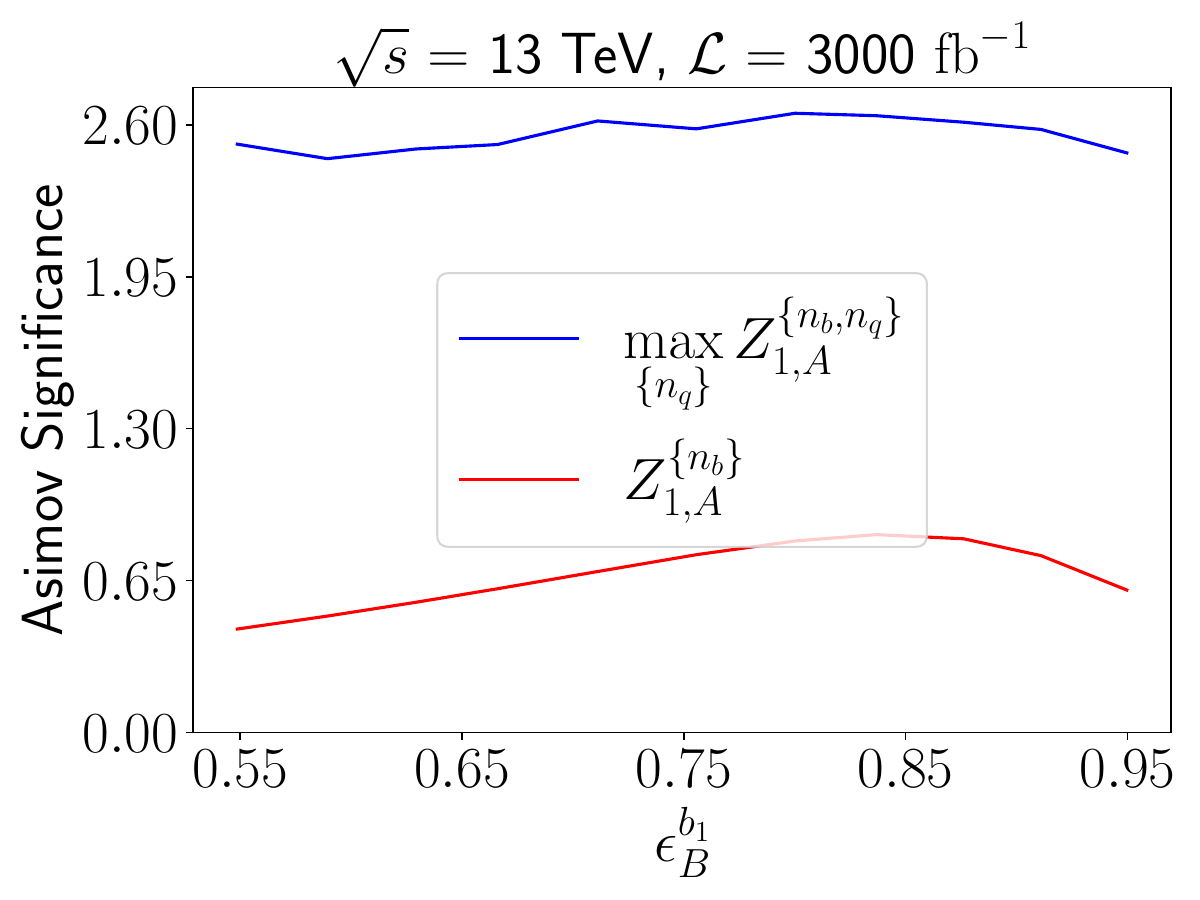}
    \end{tabular}
    \caption{Expected results for the proposed strategy using different WPs.
    Left column: expected discovery Asimov significance for the analysis that uses $\{n_{b},n_{q}\}$ bins, as a function of two $b$-tagging working point efficiencies, $\epsilon_B^{b_1}$, $\epsilon_B^{b_2}$. Right column: Expected discovery Asimov significance for analysis that uses just $\{n_{b}\}$ bins (red) compared with the highest achievable significance using $\{n_{b},n_{q}\}$ binning. Top row corresponds ot the 8 TeV LHC while the bottom row corresponds to the HL-LHC.}
    \label{fig:significance_scan_1}
\end{center}
\end{figure}

We present the results for two center of mass energies $\sqrt{s}$ and luminosities $\mathcal{L}$ in Figure ~\ref{fig:significance_scan_1}. The first row in Figure~\ref{fig:significance_scan_1} gives the sensitivity one could expect from Run 1 and should be compared with the results in Ref.~\cite{Khachatryan:2014nda}. The second row gives the sensitivity one can expect at HL-LHC (albeit using 13 TeV collision energy, instead of 14 TeV). We perform the scans using fixed tagging systematic uncertainties, whose numerical values are set to reasonable benchmarks, and which are the limiting factors for the achievable significance. Further increase in the statistical power of the analysis could be possible by a better handle of the relevant systematics~\cite{Faroughy:2022dyq}.  For each $\sqrt{s}$ and $\mathcal{L}$, we report the expected significances using $\{n_{b},n_{q}\}$ and compare, for a given $\epsilon^{b_{1}}_{B}$, the ratio between said significance and the expected significance using only $\{n_{b}\}$.

From Figure~\ref{fig:significance_scan_1}, we observe how the addition of $n_{q}$ noticeably increases the significance of the analysis for a wide range of WPs for both benchmarks. The near-horizontality of the significance evolution shows that $n_{q}$ is carrying most of the statistical power. For $\sqrt{s}=8$ TeV there is a relative increase of around 4.5 for most $\epsilon^{b_{1}}_{B}$ although the resulting significance is still very low, reflecting the fact that $\mathcal{R}^{\mathrm{SM}}_{b}$ is indistinguishable from 1 as seen in the previous section and in Ref.~\cite{Khachatryan:2014nda}. The power of the analysis increases for medium $\epsilon^{b_{2}}_{B}$. When $\epsilon^{b_{2}}_{B}$ is high enough, the loss of statistics is too much for the sample purity to compensate. 

This is no longer the case for the HL-LHC where the statistics is higher and thus higher $\epsilon^{b_{2}}_{B}$ corresponds to higher significance. The expected significance reaches maximum values of around 2.5$\sigma$, which is considerably higher than the 0.85$\sigma$ achievable with only $\{n_{b}\}$, and reflects the clear possibility of measuring $\mathcal{R}^{\mathrm{SM}}_{b}$ directly at the HL-LHC by looking at dileptonic $t\bar{t}$ production. This is achieved by obtaining the highest purity available in the $n_{q}$ bins, which forces the $\mathcal{R}_{b}=1$ hypothesis to push $\epsilon^{q}_{B}$ to higher values through its nuisance parameter resulting in a tension with the Asimov dataset.


\section{Final remarks}

In this proceeding we have detailed a proposal to extend a previous analysis strategy by incorporating additional information in the form of the number of $q$-tagged jets $n_q$. Having verified that the probabilistic model captures dileptonic top pair production events appropriately we project its sensitivity for the HL-LHC and observe that the proposed strategy allows to measure non-null $|V_{tq}|$ at the 2$\sigma$ level. However the probabilistic model is incomplete and additional nuisance parameters should be included. It could also be extended to be more physical. For example, the model could incorporate jet kinematics by conditioning on $p_T$. Finally, we should mention that although we have treated tagger efficiency estimation and $\mathcal{R}_{b}$ determination as separate problems, they are related and could be treated at the same time.

\Acknowledgements
MS acknowledges support in part by the DOE grant de-sc0011784 and NSF OAC-2103889.

\bibliography{eprint}{}
\bibliographystyle{unsrt}
 
\end{document}